\documentstyle[prl,aps,twocolumn,epsfig,floats]{revtex}

\begin{document}
\bibliographystyle{unsrt}
\draft
\title{Stable Coulomb bubbles ?}
\author{L.G. Moretto, K. Tso and G.J. Wozniak}
\address{Nuclear Science Division, Lawrence Berkeley Laboratory, Berkeley,
California 94720}

\date{\today}

\maketitle

\begin{abstract}

Coulomb bubbles, though stable against monopole displacement, are unstable at least
with respect to quadrupole and octupole distortions.  We show that there exists a
temperature at which the pressure of the vapor filling the bubble stabilizes all the radial
modes.  In extremely thin bubbles, the crispation modes become unstable due to
the surface-surface interaction.

\end{abstract}

\narrowtext

The possibility of stable or metastable non-spherical nuclear configurations, like bubbles
or tori, has been occasionally considered 
\cite{Whe50,Sie67,Wong72,Wong73,Wong77,Wong85}.  
Earlier studies, based upon the liquid drop model, showed the presence of a bubble 
monopole minimum above a certain fissility parameter (Coulomb bubble) \cite{Wong73}.  
However, the higher deformation modes of the bubble appeared to be unstable.  
A recent calculation using the generalized rotating liquid drop model has shown
the appearance of metastable bubble-like minima at high angular momentum \cite{Royer96}.
Similarly, finite temperature Hartree Fock and Thomas Fermi calculations give indications
of the onset of bubble formation \cite{Wong85}.
Recent simulations of nuclear collisions by means of transport (BUU) equations indicate
the possibility of bubble formation \cite{Bord93a,Bord93b,Baue92,Xu94}.

Coulomb bubbles, their formation, stability, and eventual demise are of broad 
interest, and are relevant not only to nuclei, but also to highly electrified fluids 
when the Coulomb interaction becomes dominant over the surface tension. 

In what follows, we will show how the vapor pressure solves the outstanding problem of the
secular stability of Coulomb bubbles.  Furthermore we shall illustrate the role of a 
recently discovered surface instability (sheet instability) 
\cite{More92} in their eventual demise.

Within the framework of the liquid drop model, the energy $E$ of a bubble in units of 
twice the surface energy of the equivalent sphere (constant volume) can be easily
written down as a function of the bubble monopole coordinate $x$:

\begin{eqnarray}
E & = & \frac{1}{2} x^2 + \frac{1}{2} \left( 1+x^3 \right)^{2/3}
 + {\rm X} \left[ \left( 1 + x^3 \right) ^{5/3} + \frac{3}{2} x^5 \right. \nonumber \\
& &  \left. - \frac{5}{2} x^3 \left( 1 + x^3 \right) ^{2/3} \right] 
 + \frac{R}{\left[ \left( 1 + x^3 \right) ^{5/3} - x^5 \right] - x^3 P }. 
\end{eqnarray}
Here $x$ is defined as the ratio of the inner sphere radius $R_1$ over the
radius of the equivalent sphere $R_o$.
The {\it Coulomb}, {\it angular momentum}, and {\it pressure} terms are defined in terms
of the fissility parameter ${\rm X}$, rotational energy $R$, and reduced pressure $P$
respectively:
\begin{eqnarray}
{\rm X} = \frac{E_c^o}{2E_s^o}, 
\quad R = \frac{E_R^o}{2E_s^o},
\quad P = \frac{pV_o}{2E_s^o}. \nonumber
\end{eqnarray}
Here the common denominator $2E_s^o$ is twice the surface energy of the equivalent sphere;
$E_c^o$ and $E_R^o$ are the Coulomb and rotational energies; $p$ and $V_o$ are the
actual pressure and equivalent sphere volume respectively.

At zero pressure and angular momentum, the surface energy increases as a bubble develops
from a sphere, but the Coulomb energy decreases as the charges are brought farther apart
due to the bubble expansion.  Therefore, an interplay between the Coulomb and surface
energies may generate a minimum energy point along the monopole coordinate.
The bubble minimum appears first at a value of the fissility parameter 
${\rm X} = 2.022$, and becomes the absolute minimum at ${\rm X} = 2.204$ 
\cite{Wong73,Kin96}.  How can such large values of ${\rm X}$ be accessible, if 
the value of ${\rm X}$ for $^{238}U$ is only 0.714, and even for the nucleus
arising from the fusion of two nuclei of $^{238}U$, ${\rm X} = 1.427$?  The obvious
possibility lies in higher temperatures, which decrease the surface energy 
coefficient (which must go to zero at the critical temperature).  For instance, within
the framework of a Thomas-Fermi calculation \cite{Guet88,Jaqa89},
a nucleus like $^{238}U + ^{238}U$ achieves the critical value ${\rm X}=2.204$ for
bubble formation at $T= 8.13$ MeV.
  
\begin{figure}[t]
\centerline{\epsfig{file=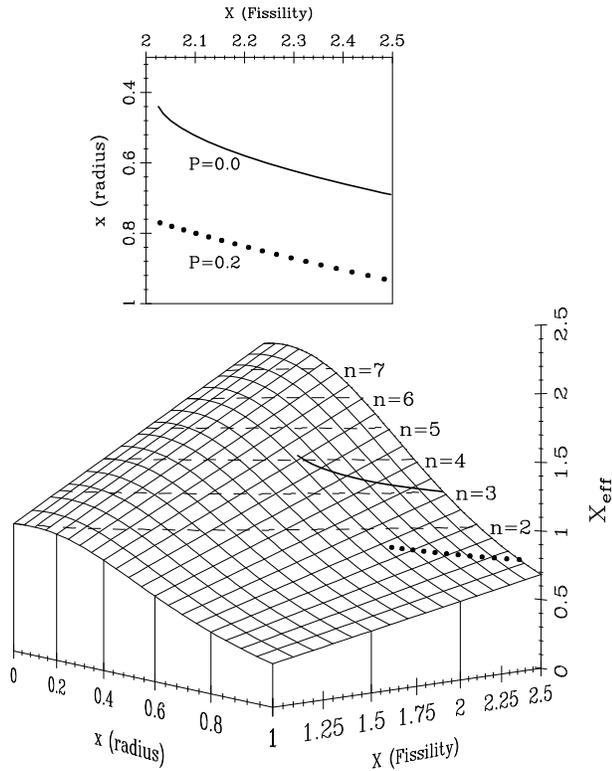,width=8.0cm,angle=180}}
\vspace*{0.4cm}
\caption{Effective fissility parameter ${\rm X_{eff}}$ as a function of the fissility
parameter ${\rm X}$ of the equivalent sphere and the inner radius $x$ of the
bubble.  The dashed lines indicate the onset of instability for specified modes.
The solid and dotted curves plot the value of ${\rm X_{eff}}$ as a function of 
${\rm X}$ for reduced pressures at 0.0 and 0.2 
respectively.  (Upper insert) The projections of the solid and dotted curves 
on the $x-{\rm X}$ plane.}
\label{Xeff_all}
\end{figure}

The solid line in the upper insert of Fig. (\ref{Xeff_all}) plots the dimensionless 
monopole coordinate of the bubble minimum as a function 
of the fissility parameter ${\rm X}$.
The radius of a Coulomb bubble is found to increase with the fissility parameter 
${\rm X}$.  The spherical minimum and the bubble minimum are separated by a barrier
whose maximum value $\Delta E_b = 0.0306E_s^o$ is attained at ${\rm X} = 2.022$.

Similarly, at zero fissility and 
pressure, there exists a critical value $(R=0.953)$ of the rotational parameter at 
which a bubble first appears, and a second critical value $(R=1.055)$
at which the bubble minimum becomes the deeper minimum.


The pressure, on the other hand, does not give rise to a bubble minimum on its own.  At
constant pressure, zero fissility, and zero angular momentum, 
the sphere minimum is the only minimum.  When $x$ increases, a barrier is
encountered beyond which there is a runaway expansion of the bubble.  At constant $Px^3$,
like at constant temperature and molar number, the pressure term becomes a constant
energy shift, and the energy rises indefinitely with $x$ like the total surface energy.

A Coulomb bubble that is stable against monopole oscillations, may be
subjected to higher order perturbations.  
The higher deformation modes of the bubble can be divided into two classes \cite{Kin96}:
%
the {\it \underline {radial modes}} and the {\it \underline {crispation modes}}.
The deformations on the two surfaces are in phase with each other for a radial mode, and
they are out of phase for a crispation mode.

The monopole oscillation obviously belongs to the class of radial modes.  On the other 
hand, the lowest order crispation mode is the dipole mode which corresponds to a rigid 
displacement of the two spheres, one with respect to the other.  Notice that this mode, 
in the absence of the Coulomb and rotational terms, is indifferent, and leads to the 
eventual puncturing of the bubble.  The introduction of the Coulomb term tends to 
stabilize a bubble against crispation dipole oscillation.  The radial dipole mode, however,
is trivial since it involves only the motion of the center-of-mass.  Hence, a nuclear
bubble is always stable with respect to a dipole perturbation within our present
description.

Unlike the dipole oscillation, higher multipole perturbations tend to increase the 
surface energy,
and thus stabilize the unperturbed bubbles.  This surface effect is the same for the
radial and crispation modes, since the two modes differ only in the relative orientation
of their surfaces.  On the other hand, the Coulomb effect is drastically different
for the two modes.  The Coulomb perturbation energy is always negative for the radial
mode, since the average distance between charges is increased slightly due to the
perturbation.  A similar effect of Coulomb destabilization is observed for the crispation
mode in case of thick bubbles.  In fact, the two modes are indistinguishable for a solid
sphere.  However, this destabilization effect becomes progressively weaker as the
bubble expands.  When a bubble is sufficiently thin, the Coulomb perturbation energy
becomes positive, and stabilizes the crispation modes.
This is because the Coulomb force tends to resist the attempt to concentrate the charge
in ``clumps'' distributed on the surface of the thin bubble, as required by the
higher order crispation modes.  In general, 
the Coulomb destabilization effect is always stronger for the radial modes.  
Therefore, a bubble that is stable with respect to radial perturbations is 
always stable against crispation perturbations within our present description.

To see the role of the Coulomb term on the stability of radial modes, let us
recall that for a charged drop, the reduced frequency of the $n^{th}$ modes is given by
\cite{More93}
\begin{equation}
\omega ^2 = \frac{1}{8} n(n-1) \left\{ (n+2) - 4 {\rm X} \right\}.
\label{charged_drop}
\end{equation}
Notice that for ${\rm X} = 1$ the frequency goes to zero for $n=2$.  This is the onset
of quadrupole instability, or the well known fission instability.  For ${\rm X} > 1$
progressively higher modes are destabilized.  The \underline{last unstable} mode is
$n_{last} = 4 {\rm X} - 2$.  For instance, $n_{last}$ increases from 10 to 14 as
${\rm X}$ is incremented from 3 to 4.  
This shows that an increase of the Coulomb force destabilizes a larger number of radial 
modes.  In addition, Eq. (\ref{charged_drop}) allows one to define the
\underline{most unstable} mode (negative minimum of $\omega^2$).
For example, the most unstable modes are 7 and 10 for ${\rm X}$ = 3 and 4, respectively.  
Hence, a highly charged sphere will not merely fission, but will break up in many 
droplets through an instability associated with a high multipole mode.
Interestingly, the most unstable mode does not coincide with $n_{last}$, nor with
the lowest (fission) mode either.


Eq. (\ref{charged_drop}) can be applied to the radial modes of the bubble as well, 
provided that, at any given value of the monopole coordinate $x$, an
``effective'' fissility parameter is defined
\begin{eqnarray}
{\rm X}_{eff} = \frac {E_c(x)}{2E_s(x)}. \nonumber
\end{eqnarray}
Since the Coulomb term decreases with $x$, while the corresponding surface term increases,
the value of ${\rm X_{eff}}$ decreases 
as the bubble expands at a given fissility parameter, as shown in
Fig. (\ref{Xeff_all}).  If the original nucleus ($x$=0) is unstable up to the multipole of
order $n$, as it develops into a bubble ($x>0$) it starts stabilizing the higher 
order radial modes.  The dashed lines in Fig. (\ref{Xeff_all}) show that the last unstable
mode decreases with increasing $x$. 


The solid curve in Fig. (\ref{Xeff_all}) indicates the values of ${\rm X_{eff}}$ associated
with the bubble minima at different fissilities.  At the threshold fissility 
of ${\rm X}$ = 2.022, the value of ${\rm X_{eff}}$ lies just about at the $n$=4 stability
line, indicating that the bubble is unstable up to the $n=4$ mode.
As more charge is brought into the bubble with increasing values of ${\rm X}$, 
the Coulomb bubble expands and it becomes stable with respect to the $n=4$ and
even to the octupole mode ($n=3$) at ${\rm X}$=2.5.  However, the
Coulomb bubble is still unstable with respect to the quadrupole mode ($n=2$).  
In fact, a further increase of ${\rm X}$ does not stabilize the quadrupole mode.



Yet, it may be possible to have a stable nuclear bubble.  If the bubble is warm,
it fills up with vapor arising from the fluid itself.  The effect of pressure on the 
stability of the radial modes is most remarkable!
The resulting pressure acts only upon the monopole mode, by displacing
outwards the Coulomb minimum.  The effect on the other radial modes is nil, since only
changes in volume are relevant to pressure.  Consequently, the positions in $x$ of the last
unstable modes for a fixed value of ${\rm X}$ do not change.
The dotted curve in the insert of Fig. (\ref{Xeff_all}) shows the expansion of the Coulomb
bubble provided by a reduced pressure of 0.2.  When this dotted curve is projected onto
the surface of ${\rm X_{eff}}$, it appears below the quadrupole stability line.
{\it This shows that the bubble has become secularly stable with respect 
to all the modes}.  

\begin{figure}[t]
\centerline{\epsfig{file=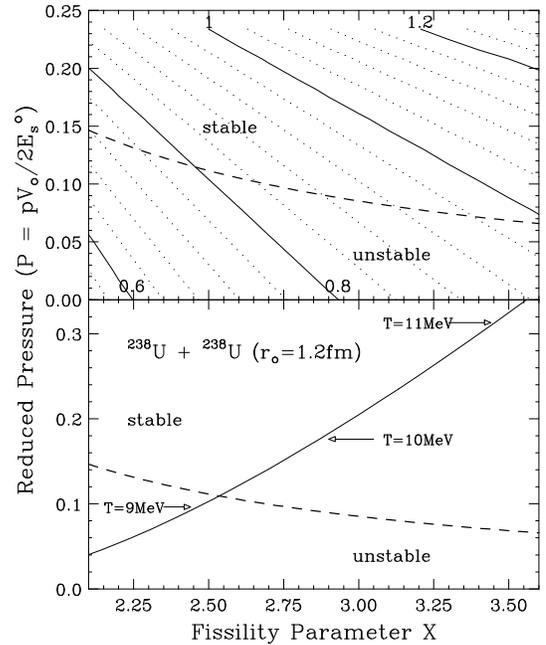,width=7.0cm,angle=180}}
\vspace*{0.4cm}
\caption{(Top) The linear contour plot (dotted and solid lines)
shows the inner radius of a bubble minimum 
as a function of reduced pressure $P$ and fissility parameter ${\rm X}$.  
The dashed line indicates the onset of instability for quadrupole mode.
(Bottom) For the system of $^{238}U + ^{238}U$, the line plots the increasing 
values of $P$ and ${\rm X}$ with temperature.  
The dashed line is the dashed line from the top panel.}
\label{bubble567}
\end{figure}

To study this pressure effect in combination with the fissility parameter, 
a contour plot indicating 
the inner radius at the bubble minimum is shown as a function of $P$ and ${\rm X}$ 
in the top panel of Fig. (\ref{bubble567}).
The lower limit of ${\rm X}$ is 2.022, the fissility at which a bubble minimum first 
appears.  The dashed line indicates the onset of instability for the quadrupole mode, which
also define the boundary conditions of bubble stability against all the radial modes. 
It can be seen that at a given value of ${\rm X}$, it is always possible to find a 
pressure large enough to shift the bubble minimum to a thinner and stable configuration.


A natural source for this pressure, in the case of nuclei or other fluids in vacuum, is 
the pressure of the saturated vapor, which spontaneously fills up the bubble if $T > 0$.  
As the outer surface is looking into vacuum, one might think that no pressure is exerted 
on it.  However, since the outer surface is constantly evaporating, an ablation pressure is
generated.  Since the average impulse brought in by a vapor particle is equal at 
equilibrium to that of the evaporated particle, it follows that the ablation pressure is
exactly equal to one half of the vapor pressure.
  
Using the Thomas-Fermi model \cite{Kupp74}, a temperature can 
always be found at which the vapor pressure stabilizes the bubble against all the radial
mode perturbations.  An example for the system of $^{238}U + ^{238}U$ is shown in
the bottom panel of Fig. (\ref{bubble567}).  The dashed line is equivalent to the dashed
line in the top panel, 
which defines the boundary conditions of bubble stability 
against all the radial modes.  The solid line shows the temperature effect on both
the reduced pressure and the fissility parameter.  In this case, a nuclear temperature
of about 10 MeV is sufficient to stabilize a bubble configuration against
perturbations of all radial modes.


Thus far, we have considered the effects of surface, charge, and pressure, on distorted
bubbles, and found that a) stability against radial perturbations can be achieved, and 
b) that it is a sufficient condition for the overall bubble stability.
However, when a bubble becomes rather thin, a possible demise of the bubble may be 
associated with the sheet instability which has not been treated here so far.
The sheet instability \cite{More92} is a new kind of Rayleigh-like
surface instability associated with the crispation modes. 
A nuclear sheet of any thickness tends to escape from the high surface energy by
breaking up into a number of spherical fragments with less overall surface.
However, any perturbation of finite wavelength increases the surface area, and consequently
the energy of the sheet, independent of the sheet thickness.  Clearly, this
barrier prevents the sheet from reaching the more stable configurations.  However,
when a nuclear sheet becomes sufficiently thin, the two nuclear surfaces interact with
each other.  This proximity interaction may become sufficiently strong to overcome 
the sharp barrier and causes the sheet to puncture into numerous fragments.
Using the expression in Ref. \cite{Bloc77} for the proximity potential, a critical
wavelength is determined for the onset of this surface instability for a flat sheet:
$\lambda _c = 1.1 b \cdot exp(2d/3b)$,
where $b$ is the range of the proximity interaction and $d$ is the thickness of the sheet.

A bubble behaves much like a sheet, and is subject to the sheet instability.  Since a
bubble, like a sheet, must rely on the proximity interaction to become unstable, it will
retain its surface stability until the range of the surface-surface interaction
is of the order of its thickness.  Thus a critical range of proximity interaction for
the onset of bubble instability against crispation perturbation can be defined as
$b_c = f \left( x, {\rm X}, n \right)$.

\begin{figure}[t]
\centerline{\epsfig{file=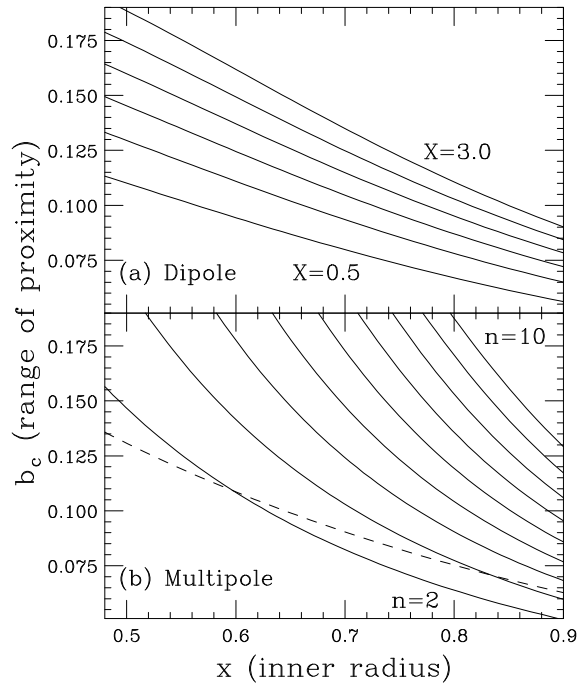,height=9.0cm,angle=180}}
\vspace*{0.4cm}
\caption{(a) Critical range of proximity interaction ($b_c$) as a function of inner radius
($x$) for the dipole mode at various fissility parameters (${\rm X}=0.5-3.0$).  
(b) $b_c$ as a function of $x$ for multipole modes ($n=2-10$) of a neutral bubble.  The
dashed line indicates values of $b_c$ for a charged bubble (${\rm X}=1.5$) undergoing
quadrupole perturbation.}
\label{crispation}
\end{figure}

Fig. (\ref{crispation}a) plots the value of $b_c$ for the onset of dipole instability at 
the indicated values of fissility.  Notice that the line for ${\rm X}=0$ is missing,
since the dipole mode of a neutral bubble is indifferent, and any finite proximity 
effect is sufficient to trigger the instability.  Recall that the introduction of charge
stabilizes a bubble against the dipole oscillation, and thus offsets the proximity effect.
Consequently, the value of $b_c$ at any given bubble radius increases with ${\rm X}$
as shown in Fig. (\ref{crispation}a).

Unlike the dipole mode, the surface energy of higher multipole perturbations increases
monotonically with the bubble radius.  To study the interplay between this surface effect
and the proximity interaction, a neutral bubble is considered.  The solid lines in
Fig. (\ref{crispation}b) plot $b_c$ as a function of $x$ for progressively higher
order modes ($n=2 - 10$).  
Clearly, the quadrupole instability is most easily triggered 
among the multipole modes.  As the proximity interaction becomes 
stronger (larger $b_c$), higher order multipoles are gradually destabilized.
The dashed line in Fig. (\ref{crispation}b) shows the onset of quadrupole instability
for a charged bubble with ${\rm X}=1.5$.  Interestingly, the dashed and the corresponding
solid lines cross at about $x=0.6$, reflecting different Coulomb effects mentioned
earlier for thin and thick bubbles undergoing multipole crispation perturbations.
An increase in charge stabilizes a bubble against 
higher order modes and offsets the proximity effect (larger $b_c$) until it becomes 
sufficiently thick ($x < 0.6$ for the quadrupole mode at ${\rm X}=1.5$).

In conclusion, the depletion of charge in the central cavity of nuclear bubbles reduces
the Coulomb energy significantly and thus stabilizes ``Coulomb'' bubbles against monopole
oscillations.  These Coulomb bubbles, however, are at least unstable to
perturbation of the quadrupole radial mode.  On the other hand, 
a sufficiently high temperature generates a vapor pressure in the central cavity
which drives the bubble to a thinner configuration that is
stable against all the radial modes.
Finally, a thin Coulomb bubble behaves like a sheet, and becomes
susceptible to a proximity surface instability via the crispation modes when its 
thickness is comparable to the range of the proximity interaction.


This work was supported by the Director, Office of Energy Research, Division of Nuclear
Physics of the Office of High Energy and Nuclear Physics of the U.S. Department of Energy
under Contract No. DE-AC03-76SF00098.


\vspace*{-0.2cm}
\begin{thebibliography}{99}
\vspace*{-1.2cm}
\bibitem{Whe50} J.A. Wheeler, Nucleonics Notebook, unpublished, (1950).
\bibitem{Sie67} P. Siemens and H. Bethe, Phys. Rev. Lett. {\bf 18}, 704 (1967).
\bibitem{Wong72} C.Y. Wong, Phys. Lett. {\bf B41}, 446 (1972); {\bf B41}, 451 (1972).
\bibitem{Wong73} C.Y. Wong, Ann. Phys. {\bf 77} 279 (1973).
\bibitem{Wong77} C.Y. Wong {\it et al.}, Phys. Lett. {\bf B66}, 19 (1977).
\bibitem{Wong85} C.Y. Wong, Phys. Rev. Lett. {\bf 55}, 1973 (1985).
\bibitem{Royer96} G. Royer {\it et al.}, Nucl. Phys. {\bf A605}, 403 (1996).
\bibitem{Bord93a} B. Borderie {\it et al.}, Phys. Lett. {\bf B302}, 15 (1993).
\bibitem{Bord93b} B. Borderie {\it et al.}, Phys. Lett. {\bf B307}, 404 (1993).
\bibitem{Baue92} W. Bauer {\it et al.}, Phys. Rev. Lett. {\bf 69}, 1888 (1992).
\bibitem{Xu94} H.M. Xu {\it et al.}, Phys. Rev. {\bf C49}, 1778 (1994).
\bibitem{More92} L. G. Moretto {\it et al.}, Phys. Rev. Lett. {\bf 69}, 1884 (1992).
\bibitem{Kin96} K. Tso, Ph.D. Thesis, LBL 38884 (1996).
\bibitem{Guet88} C. Guet {\it et al.}, Phys. Lett. {\bf B205}, 427 (1988).
\bibitem{Jaqa89} H.R. Jaqaman, Phys. Rev. {\bf C40}, 1677 (1989).
\bibitem{More93} L. G. Moretto and G. J. Wozniak, Ann. Rev. Part. \& Nucl. Sci ,  
{\bf Vol. 43}, 379 (1993).
\bibitem{Kupp74} Wolfgang A. Kupper {\it et al.}, Ann. Phys. {\bf 88}, 454 (1974).
\bibitem{Bloc77} J. Blocki {\it et al.}, Ann. Phys. {\bf 105}, 427 (1977).
\end{thebibliography}
\end{document}